\documentclass[12pt]{article}
\usepackage{amsmath}
\usepackage{amssymb}
\usepackage{epsfig}
\usepackage{euscript}
\def\mathswitchr#1{\relax\ifmmode{\mathrm{#1}}\else$\mathrm{#1}$\fi}

\def\rQCED{{\rm QCED}}

\newcommand {\pslash}{\hbox{$\not\hbox{\kern-2.3pt $p$}$}}

\newcommand{\FYFS}{F_{\mathrm{YFS}}}
\usepackage{cite}

\usepackage{epic}

\def\alf1{ {\alpha\over\pi} }
\def\rQCED{{\rm QCED}}
\begin{document}
\begin{titlepage}
\begin{center}
{\bf \large Phenomenological Study of the Interplay between IR-Improved DGLAP-CS Theory and the Precision of an NLO ME Matched Parton Shower MC}\\
\vspace{2mm}
    S.K. Majhi
 \footnote{Work supported by 
grant Pool No. 8545-A, CSIR, IN.}\\
      Indian Association for the Cultivation of Science, Kolkata, India\\
        E-mail: tpskm@iacs.res.in\\
A. Mukhopadhyay\\%
      Baylor University, Waco, TX, USA\\
        E-mail: aditi\_mukhopadhyay@baylor.edu\\
B.F.L. Ward\\%
      Baylor University, Waco, TX, USA\\
        E-mail: bfl\_ward@baylor.edu\\
S.A. Yost
 \footnote{Work supported in part by U.S.
D.o.E. grant DE-FG02-10ER41694 and grants from The Citadel Foundation.}\\
      The Citadel, Charleston, SC, USA\\
        E-mail: scott.yost@citadel.edu\\
\end{center}
\vspace{2mm}
\centerline{\bf Abstract}
We present a phenomenological study of the current status of the application of our approach of {\it exact} amplitude-based resummation in quantum field theory to precision QCD calculations, by realistic MC event generator methods, as needed for precision LHC physics. We discuss recent results as they relate to the interplay of the attendant IR-Improved DGLAP-CS theory of one of us and the precision of exact NLO matrix-element matched parton shower MC's in the Herwig6.5 environment as determined by comparison to recent
LHC experimental observations on single heavy gauge boson production and decay. The level of agreement between the new theory and the data continues to be a reason for optimism. In the spirit of completeness, we discuss as well other approaches to the same theoretical predictions that we make here from the standpoint of physical precision with an eye toward the (sub-)1\% QCD$\otimes$EW 
total theoretical precision regime for LHC physics. \\
\vspace{1cm}
\begin{center}
 BU-HEPP-13-02, Apr., 2013\\
\end{center}

\end{titlepage}
%

 
\def\Kmax{K_{\rm max}}\def\ieps{{i\epsilon}}\def\rQCD{{\rm QCD}}

\section{\bf Introduction}\par

With the recent announcement~\cite{atlas-cms-2012} 
of an Brout-Englert-Higgs (BEH)~\cite{EBH} candidate 
boson after the start-up and successful running 
of the LHC for three years, we have entered the era of precision QCD, 
by which we mean
predictions for QCD processes at the total precision tag of $1\%$ or better. 
Given the expected role of precision comparison between theory and experiment in determining the detailed properties of the newly discovered EBH boson candidate, the attendant need for exact, amplitude-based 
resummation of large higher order effects
is now more paramount. We have argued in Refs.~\cite{radcor2011,qced} that
such resummation allows one to have better than 1\% theoretical precision 
as a realistic goal in such comparisons, so that one can indeed 
distinguish new physics(NP) from higher order SM processes and can distinguish 
different models of new physics from one another as well. 
In what follows, with an eye toward its interplay with NLO exact, matrix element matched parton shower MC precision issues, we present the status of this approach to precision QCD for the LHC in connection with its attendant IR-improved DGLAP-CS~\cite{dglap,cs} theory~\cite{irdglap1,irdglap2} realization via HERWIRI1.031~\cite{herwiri} 
in the HERWIG6.5~\cite{herwig} environment. 
We realize here the attendant exact, NLO matrix element matched parton shower MC
realizations for both HERWIRI1.031 and HERWIG6.5 in the MC@NLO~\cite{mcatnlo} methodology, wherewith we make the corresponding
comparisons with recent LHC data that we present herein.
\par
In this way, we continue the strategy of building on existing platforms to develop and realize a path toward precision QCD for the physics of the LHC. What we exhibit is an explicit union of the new IR-improved DGLAP-CS theory and the MC@NLO realization of exact NLO matrix element(ME) matched parton shower MC theory. We note that we are also pursuing the implementation~\cite{elswh} of the new IR-improved 
DGLAP-CS theory for
HERWIG++~\cite{hwg++}, HERWIRI++,
for PYTHIA8~\cite{pyth8} and for SHERPA~\cite{shrpa}, as well as
the corresponding NLO ME/parton shower matching realizations in the POWHEG~\cite{pwhg} framework. For, our ultimate goal is a provable precision tag on our theoretical predictions and one of the strongest cross checks on theoretical precision is the difference between two independent realizations of the attendant theoretical calculation. We will present such cross checks elsewhere~\cite{elswh}.
\par
The paper is organized as follows.
We motivate the development of the new IR-improved DGLAP-CS theory
in our approach to precision QCD theory by showing
in the next section how it arises naturally in the effort to 
realize a provable precision from our approach~\cite{qced} to 
precision LHC physics. This will allow us 
to expose properly the interplay between the NLO ME matched parton shower MC precision and the new IR-improved DGLAP-CS theory. 
In the interest of completeness, we also review in the next section the relevant aspects
of the approach in Ref.~\cite{qced}, 
which is an amplitude-based QED$\otimes$QCD($\equiv\text{QCD}\otimes\text{QED}$) exact resummation 
theory 
realized by MC methods. This paper is an extension
of a shorter report of our results and analysis which is given
in Ref.~\cite{plb2013}. Accordingly, also in the interest of completeness, we discuss as well in the next section other possible approaches to the predictions we make herein from the standpoint of their physical precision. 
We believe this discussion will aid the reader in putting the results we present in a proper context.
Section 3 contains the applications to the recent LHC data on single heavy gauge boson production with an eye toward the analysis in Refs.~\cite{herwiri} of the analogous processes at the Tevatron. In this paper we will focus on the single Z/$\gamma*$ production and decay to lepton pairs for definiteness. The other heavy gauge boson processes will be taken up elsewhere~\cite{elswh}. Section 4 contains our concluding observations and our outlook.
\par

\section{Review of Our Approach to Precision LHC Physics}

We take the starting point for what we discuss here as the following
fully differential representation of a hard LHC scattering process:
\begin{equation}
d\sigma =\sum_{i,j}\int dx_1dx_2F_i(x_1)F_j(x_2)d\hat\sigma_{\text{res}}(x_1x_2s),
\label{bscfrla}
\end{equation}
where the $\{F_j\}$ and 
$d\hat\sigma_{\text{res}}$ are the respective parton densities and 
reduced hard differential cross section and where 
we use the subscript on the latter to indicate that it 
has been resummed
for all large EW and QCD higher order corrections in a manner consistent
with achieving a total precision tag of 1\% or better for the total 
theoretical precision of (\ref{bscfrla}). The key theoretical issue
for precision
QCD for the LHC is then 
The proof of the correctness of the value of the 
total theoretical precision $\Delta\sigma_{\text{th}}$ of (\ref{bscfrla})
is then the key theoretical issue for the realization of precision QCD for the
LHC. 
The theoretical precision in question can be represented as follows:
\begin{equation}
\Delta\sigma_{\text{th}}= \Delta F \oplus\Delta\hat\sigma_{\text{res}}
\label{eqdecomp1}
\end{equation}
where $\Delta A$ is the contribution of the uncertainty
on the quantity $A$ to $\Delta\sigma_{\text{th}}$. Here, we stress that we discuss the situation in which the two errors in (\ref{eqdecomp1}) are independent
for definiteness; (\ref{eqdecomp1}) has to be modified accordingly when
they are not. We further stress that the
proof of the correctness of the value of the 
total theoretical precision $\Delta\sigma_{\text{th}}$ is indispensible
in order to validate the  application of a given 
theoretical prediction to precision 
experimental observations, for the discussion of the signals and the
backgrounds for 
both Standard Model(SM) and new physics (NP) studies, and more specifically
for the overall normalization
of the cross sections in such studies.
If one uses a calculation
with an unknown value of $\Delta\sigma_{\text{th}}$ 
for the attendant studies, the NP itself can be missed.
We do feel a need to continue to state here that 
this point simply cannot be emphasized too much.\par
In the interest of completeness, 
we note that, by our definition, which follows the discussion in Ref.~\cite{jadach-prec}, $\Delta\sigma_{\text{th}}$ is the 
total theoretical uncertainty that derives from the physical and technical
precision 
contributions:
the physical precision contribution, $\Delta\sigma^{\text{phys}}_{\text{th}}$,
arises from such sources as missing graphs, approximations to graphs, 
truncations,....; the technical precision contribution, 
$\Delta\sigma^{\text{tech}}_{\text{th}}$, arises from such sources as 
bugs in codes, numerical rounding errors,
convergence issues, etc. We want to observe that here, when we reference
bugs in codes we have in mind that all gross errors such as those that give obviously wrong results, as determined by cross checks, are eliminated and we have left programming errors such as those in the logic: suppose for programming error reasons a DO-loop ends at 999 steps instead of the intended 1000 steps, resulting in a per mille level error, that could alternate in sign from event to event. As per mille level accuracy is good enough in many applications, the program would remain reliable, but it would have what we call a technical precision error at the per mille level. With this understanding, 
the total theoretical error is then 
given by
\begin{equation}
\Delta\sigma_{\text{th}}=\Delta\sigma^{\text{phys}}_{\text{th}}\oplus \Delta\sigma^{\text{tech}}_{\text{th}}.
\end{equation}
Although the desired value for $\Delta\sigma_{\text{th}}$ 
depends on the  specific
requirements of the observations, as a general rule, it should fulfill
$\Delta\sigma_{\text{th}}\leq f\Delta\sigma_{\text{expt}}$, 
where $\Delta\sigma_{\text{expt}}$ is the respective experimental error
and $f\lesssim \frac{1}{2}$, so that
the theoretical uncertainty does not significantly adversely affect the 
attendant physics analysis of the data.
\par
In the interest of completeness, we will now recapitulate the
theory we have developed to realize such precision rigorously. This theory
is the $\text{QCD}\otimes\text{QED}$ resummation theory in Refs.~\cite{qced}
for the reduced cross section in (\ref{bscfrla}) and for the
resummation of the evolution of the parton densities therein as well.
More precisely, the master formula, which applies both to the resummation of the reduced cross section
and to that of the evolution of the parton densities,
may be identified as
\begin{eqnarray}
&d\bar\sigma_{\rm res} = e^{\rm SUM_{IR}(QCED)}
   \sum_{{n,m}=0}^\infty\frac{1}{n!m!}\int\prod_{j_1=1}^n\frac{d^3k_{j_1}}{k_{j_1}} \cr
&\prod_{j_2=1}^m\frac{d^3{k'}_{j_2}}{{k'}_{j_2}}
\int\frac{d^4y}{(2\pi)^4}e^{iy\cdot(p_1+q_1-p_2-q_2-\sum k_{j_1}-\sum {k'}_{j_2})+
D_\rQCED} \cr
&\tilde{\bar\beta}_{n,m}(k_1,\ldots,k_n;k'_1,\ldots,k'_m)\frac{d^3p_2}{p_2^{\,0}}\frac{d^3q_2}{q_2^{\,0}},
\label{subp15b}
\end{eqnarray}\noindent
where $d\bar\sigma_{\rm res}$ is either the reduced cross section
$d\hat\sigma_{\rm res}$ or the differential rate associated to a
DGLAP-CS~\cite{dglap,cs} kernel involved in the evolution of the $\{F_j\}$ and 
where the {\em new} (YFS-style~\cite{yfs,yfs-jw}) {\em non-Abelian} residuals 
$\tilde{\bar\beta}_{n,m}(k_1,\ldots,k_n;k'_1,\ldots,k'_m)$ have $n$ hard gluons and $m$ hard photons and we show the final state with two hard final
partons with momenta $p_2,\; q_2$ specified for a generic $2f$ final state for
definiteness. The infrared functions ${\rm SUM_{IR}(QCED)},\; D_\rQCED\; $
are defined in Refs.~\cite{qced,irdglap1,irdglap2} as follows:
\begin{eqnarray}
{\rm SUM_{IR}(QCED)}=2\alpha_s\Re B^{nls}_{QCED}+2\alpha_s{\tilde B}^{nls}_{QCED}\cr
D_\rQCED=\int \frac{d^3k}{k^0}\left(e^{-iky}-\theta(K_{max}-k^0)\right){\tilde S}^{nls}_{QCED}
\label{irfns}
\end{eqnarray}
where the dummy parameter $K_{max}$ is such that nothing depends on it and where we have introduced
\begin{eqnarray}
B^{nls}_{QCED} \equiv B^{nls}_{QCD}+B^{nls}_{QED},\cr
{\tilde B}^{nls}_{QCED}\equiv {\tilde B}^{nls}_{QCD}+{\tilde B}^{nls}_{QED}, \cr
{\tilde S}^{nls}_{QCED}\equiv {\tilde S}^{nls}_{QCD}+{\tilde S}^{nls}_{QED}.
\end{eqnarray} 
Here, the superscript $nls$ denotes that the infrared functions are DGLAP-CS synthesized as explained in Refs.~\cite{dglpsyn,qced,irdglap1,irdglap2} and the infrared functions
$B_A,\; {\tilde B}_A,\; {\tilde S}_A, \; A=QCD,\; QED,$ are given
in Refs.~\cite{yfs,yfs-jw,qced,irdglap1,irdglap2}. 
We stress that  
simultaneous resummation of QED and QCD large IR effects is exact here.\par 
We may describe the physical meanings of the key components in the master formula (\ref{subp15b}) as follows. 
In the language of Ref.~\cite{gatheral}, 
the exponent ${\rm SUM_{IR}(QCED)}$ sums up to the infinite  order the maximal leading IR singular terms in the cross section for soft emission 
below a dummy parameter $K_{\text{max}}$ and the exponent
$D_\rQCED$ does the same for the regime above $K_{\text{max}}$ so that
(\ref{subp15b}) is independent of $K_{\text{max}}$ -- it cancels between
${\rm SUM_{IR}(QCED)}$ and $D_\rQCED$\footnote{If we want to include more of the
maximal exponentiating terms from the formalism of Ref.~\cite{gatheral} in
the two exponents ${\rm SUM_{IR}(QCED)},\;D_\rQCED$, we may do so with a consequent change in the attendant residuals $\tilde{\bar\beta}_{n,m}$.}.
In order to maintain exactness order by order in perturbation theory in both 
$\alpha$ and $\alpha_s$ in the presence of these resummed terms, as explained
in Refs.~\cite{qced,irdglap1,irdglap2}
we generate the residuals $\tilde{\bar\beta}_{n,m}$ by iterative 
computation to match the attendant 
exact results to all orders in $\alpha$ and $\alpha_s$. We need to emphasize
the following most important point. In our formulation in (\ref{subp15b})
{\it the entire soft gluon phase space is included in the representation -- no part of it 
is dropped}.\par
As it is explained in Refs.~\cite{qced}, 
the new non-Abelian residuals $\tilde{\bar\beta}_{m,n}$ 
allow rigorous shower/ME matching via their shower subtracted analogs:
in (\ref{subp15b}) we make the replacements
\begin{equation}
\tilde{\bar\beta}_{n,m}\rightarrow \hat{\tilde{\bar\beta}}_{n,m}
\end{equation}
where the $\hat{\tilde{\bar\beta}}_{n,m}$ have had all effects in the showers
associated to the $\{F_j\}$ removed from them. To see how we make
contact between the $\hat{\tilde{\bar\beta}}_{n,m}$ and the
differential distributions in MC@NLO we proceed as
follows. We represent the MC@NLO differential cross section 
via~\cite{mcatnlo} 
\begin{equation}
\begin{split}
d\sigma_{MC@NLO}&=\left[B+V+\int(R_{MC}-C)d\Phi_R\right]d\Phi_B[\Delta_{MC}(0)+\int(R_{MC}/B)\Delta_{MC}(k_T)d\Phi_R]\nonumber\\
&\qquad\qquad +(R-R_{MC})\Delta_{MC}(k_T)d\Phi_Bd\Phi_R
\label{mcatnlo1}
\end{split}
\end{equation}
where $B$ is Born distribution, $V$ is the regularized virtual contribution,
$C$ is the corresponding counter-term required at exact NLO, $R$ is the respective
exact real emission distribution for exact NLO, $R_{MC}=R_{MC}(P_{AB})$ is the parton shower real emission distribution
so that the Sudakov form factor is 
$$\Delta_{MC}(p_T)=e^{[-\int d\Phi_R \frac{R_{MC}(\Phi_B,\Phi_R)}{B}\theta(k_T(\Phi_B,\Phi_R)-p_T)]}$$,
where as usual it describes the respective no-emission probability.
The respective Born and real emission differential phase spaces are denoted by $d\Phi_A, \; A=B,\; R$, respectively.
We may note further that the representation of the differential distribution
for MC@NLO in (\ref{mcatnlo1}) is an explicit realization of the compensation 
between real and virtual divergent soft effects discussed in the 
Appendices of Refs.~\cite{irdglap1,irdglap2} in establishing the validity of 
(\ref{subp15b}) for QCD -- all of the terms on the RHS of (\ref{mcatnlo1}) are 
infrared finite. Indeed,
from comparison with (\ref{subp15b}) restricted to its QCD aspect we get the identifications, accurate to ${\cal O}(\alpha_s)$,
\begin{equation}
\begin{split}
\frac{1}{2}\hat{\tilde{\bar\beta}}_{0,0}&= \bar{B}+(\bar{B}/\Delta_{MC}(0))\int(R_{MC}/B)\Delta_{MC}(k_T)d\Phi_R\\
\frac{1}{2}\hat{\tilde{\bar\beta}}_{1,0}&= R-R_{MC}-B\tilde{S}_{QCD}
\label{eq-mcnlo}
\end{split}
\end{equation}
where we defined~\cite{mcatnlo} $$\bar{B}=B(1-2\alpha_s\Re{B_{QCD}})+V+\int(R_{MC}-C)d\Phi_R$$ and we understand here
that the DGLAP-CS kernels in $R_{MC}$ are to be taken as the IR-improved ones
as we exhibit below~\cite{irdglap1,irdglap2}. 
Here we have written the QCD virtual and real infrared functions
$B_{QCD}$ and $\tilde{S}_{QCD}$ respectively without the superscript $nls$
for simplicity of notation and they are understood to be DGLAP-CS synthesized as explained in Refs.~\cite{qced,irdglap1,irdglap2} so that we
avoid doubling counting of effects. We also re-emphasize that we do not drop
any effects here in (\ref{eq-mcnlo}). We observe further that, in view of 
(\ref{eq-mcnlo}), 
the way to the extension of frameworks such as MC@NLO to exact higher
orders in $\{\alpha_s,\;\alpha\}$ is therefore open via our $\hat{\tilde{\bar\beta}}_{n,m}$
and will be taken up elsewhere~\cite{elswh}.
\par
We point out that in Refs.~\cite{irdglap1,irdglap2,herwiri} our methods for resummation of the QCD theory have been shown to be
fully consistent with the methods in Refs.~\cite{stercattrent1,scet1}. 
Specifically, it is shown in Refs.~\cite{irdglap1,irdglap2,herwiri}
that the methods in Refs.~\cite{stercattrent1,scet1}
give approximations to our hard gluon residuals  $\hat{\tilde{\bar\beta}}_{n,0}$;
 for, unlike the master formula in (\ref{subp15b}), the methods in Refs.~\cite{stercattrent1,scet1} are not exact results. To see this, observe that the threshold-resummation
methods in Refs.~\cite{stercattrent1}, using the result
that, for any function $f(z)$,
$$\left|\int_0^1 dz z^{n-1}f(z)\right|\le(\frac{1}{n})\max_{z\in [0,1]} {|f(z)|},$$
drop non-singular contributions to the cross section at $z\rightarrow 1$
in resumming the logs in $n$-Mellin space. The SCET theory in Refs.~\cite{scet1}
drops terms of ${\cal O}(\lambda)$ at the level of the amplitude, where $\lambda=\sqrt{\Lambda/Q}$ for a process with the hard scale $Q$ with $\Lambda \sim .3\text{GeV}$ so that, for $Q\sim 100\text{GeV}$, we have $\lambda\cong 5.5\%$.
From the known equivalence of the two approaches, the errors in the threshold resummation must be similar. Evidently, we can only use these approaches as a guide to our  new non-Abelian residuals as we develop results for the (sub-)1\% precision regime.\par
In view of the specific processes which we consider in the next Section,
we will continue here with our consistency discussion as it 
relates to the theory of 
QCD resummation for the specific heavy gauge boson production-type processes
offered in Refs.~\cite{colsop,colsopster}, which is again an approximate formalism that could be used to make approximations to our hard gluon residuals $\hat{\tilde{\bar\beta}}_{n,0}$. The theory in Refs.~\cite{colsop,colsopster} is in  wide use at the LHC and in the data analyses for the Tevatron -- see for example the recent analyses in Refs.~\cite{atlas-z-pt-2} where this theory, as it is implemented in the MC integration program
R{\scriptsize ES}B{\scriptsize OS}~\cite{resbos1,resbos2,resbos3} is compared to recent LHC data and to recent analyses of Tevatron data. We need to note that the theory in Refs.~\cite{colsop,colsopster} builds on and extends beyond the considerable literature in Refs.~\cite{ddt-res-hvy-bsn} aimed at the analogous processes to those under study here.
Could we perhaps employ this formalism in Refs.~\cite{colsop,colsopster} as well to reach 
$\le 1\%$ physical precision QCD predictions? Let us note that the authors
in Ref.~\cite{colsopster} give us a hint to the answer to our question in their footnote on the journal page with number 215 for that paper, wherein they equate as 'negligible' 20\% in 
discussing
possible nonperturbative contributions in their formalism. 
Let us keep this footnote in mind in what follows.\par
We first note that a defining
formula for the approach in Refs.~\cite{colsop,colsopster} is that for the differential cross section for the $p_T$ distribution for the heavy gauge boson production in hadron-hadron collisions, which we specialize here to the case of the Drell-Yan $\gamma*$ production for definiteness\footnote{The analogous results for the $W^\pm$ and $Z/\gamma*$ are obtained by straightforward substitutional manipulations of the EW aspects of the formula we record here as described in Ref.~\cite{colsopster} so that we omit such manipulations here without loss of content of the QCD aspects
our discussion.}, where we record the result in Eq.(2.2) from
Ref.~\cite{colsopster}:{\small
\begin{equation}
\frac{d\sigma}{dQ^2dydQ_T^2}\sim \frac{4\pi^2\alpha^2}{9Q^2s}\left\{\int \frac{d^2b}{(2\pi)^2} e^{i\vec{Q}_T\cdot\vec{b}}\widetilde{W}(b;Q,x_A,x_B)+\; Y(Q_T;Q,x_A,x_B)\right\}
\label{css1} 
\end{equation}}
where we have the usual kinematics so that $\vec{Q}_T=\vec{p}_T$ is the $\gamma*$ transverse momentum, A,B are protons at the LHC, $s$ is the cms squared energy of the protons, $ Q^{\mu}$ is the $\gamma*$ 4-momentum so that $Q^2$ is the $\gamma*$ mass squared, and $y=\frac{1}{2}\ln(Q^+/Q^-)$ is the $\gamma*$ rapidity so that $x_A=e^yQ/\sqrt{s}$
and $x_B=e^{-y}Q/\sqrt{s}$. We have in mind that $Q$ is near $M_Z$ here. In (\ref{css1}), the term involving $\widetilde{W}$
carries the effects from QCD resummation as developed in Refs.~\cite{colsop,colsopster}
and the $Y$ term includes those contributions which are 'regular' at $Q_T=p_T \rightarrow 0$ in the sense
of Refs.~\cite{colsop,colsopster}, i.e., order by order in perturbation 
theory they are 
derived from the parts of the attendant hard scattering coefficients that are 
less singular than $Q_T^{-2}\times (\text{logs or}\; 1)$ or 
$\delta(\vec{Q_T})$ as $Q_T=p_T \rightarrow 0$. Our question concerns 
the physical precision of 
the $\widetilde{W}$ term; for, the $Y$ term is perturbative and can be computed
in principle to the required accuracy by the standard methods.\par
The result for $\widetilde{W}$ given in Refs.~\cite{colsop,colsopster} is 
as follows. When we have $b<<1/\Lambda$ where $\Lambda$ is a typical 
QCD hadronic mass scale such as the inverse of the proton radius, the 
result for $\widetilde{W}$ is~\cite{colsop,colsopster}
\begin{align}
\widetilde{W}(b;Q,x_A,x_B) &= \sum_{a,b}\int_{x_A}^1d\xi_A\int_{x_B}^1d\xi_BF_{a/A}(\xi_A;\mu)F_{b/B}(\xi_B;\mu)e^{-S_{sc}}\nonumber\\
   &\times\sum_je_j^2C_{ja}(x_A/\xi_A,b;C_1/C_2;g(\mu),\mu)C_{jb}(x_B/\xi_B,b;C_1/C_2;g(\mu),\mu),
\label{css2}
\end{align}
where we define~\cite{colsop,colsopster} 
\begin{equation}
S_{sc}=\int_{C_1^2/b^2}^{C_2^2Q^2}\frac{d\bar{\mu}^2}{\bar{\mu}^2}\large[\ln(\frac{C_2^2Q^2}{\bar{\mu}^2})A(g(\bar{\mu});C_1)+B(g(\bar{\mu});C_1,C_2)\large]
\end{equation} 
Here we show explicitly the dependences of the parton density functions $\{F_\ell\}$ (note that our $F_\ell(x)$ corresponds to $f_\ell(x)/x$ in Refs.~\cite{colsop,colsopster}), and we show as well the dependences of the perturbatively calculable
exponentiation and scattering 
coefficient functions
$A,\; B,\;$ and $C$ as they are given in Refs.~\cite{colsop,colsopster}. The scale $\mu$ is usually set to $C_1/b$
and $C_1$ and $C_2$ are order 1 constants chosen to optimize the resultant perturbation expansions -- see Refs.~\cite{colsop,colsopster} for more discussion on this point.
The authors in Ref~\cite{colsop,colsopster} note that this result for $\widetilde{W}(b)$, when
$1/Q<< b << 1/\Lambda$, is accurate up to terms ${\cal O}(m_q/Q,\; 1/(bQ))$ where
$m_q$ represents the quark masses. In addition, the latter authors argue that
in the result (\ref{css1}) they have dropped terms of ${\cal O}(m_q/Q,\; Q_T/Q)$
in the regime where $0\le Q_T\ll Q$. These last two statements are
seen explicitly in Sect. 9 of Ref.~\cite{colsop} in eq.(9.1) and the 
equation immediately preceeding it, for reference.
\par
When we have to consider as well the regime
$b\gtrsim 1/\Lambda$, the authors in Ref.~\cite{colsop,colsopster} argue that we have to replace the functions $A,\; B,\;$ and $C$ according to 
\begin{align}
&A(g(\bar{\mu});C_1)\rightarrow A(g(\bar{\mu}),m_q/\bar{\mu};C_1)\nonumber\\
&B(g(\bar{\mu});C_1,C_2)\rightarrow B(g(\bar{\mu}),m_q/\bar{\mu};C_1,C_2)\nonumber\\
&\sum_{a,b}\int_{x_A}^1d\xi_A\int_{x_B}^1d\xi_BF_{a/A}(\xi_A;\mu)F_{b/B}(\xi_B;\mu)\sum_je_j^2C_{ja}(x_A/\xi_A,b;C_1/C_2;g(\mu),\mu)\nonumber\\
&\times C_{jb}(x_B/\xi_B,b;C_1/C_2;g(\mu),\mu)\rightarrow \sum_je_j^2\bar{{\cal P}}_{j/A}(x_A,b;C_1/C_2)\bar{{\cal P}}_{j/B}(x_B,b;C_1/C_2)
\label{css3}
\end{align}
where we see that we can no longer neglect the quark masses and that
we no longer have the convolutions $C*F$: the functions $\bar{{\cal P}}_{j/A}$
reduce to the latter form when we take $b << 1/\Lambda$ and drop terms ${\cal O}(b\Lambda)$, according to the arguments in Refs.~\cite{colsop,colsopster}.\par
The authors in Refs.~\cite{colsop,colsopster} then join these two results, that for the regime $b << 1/\Lambda$ and that for the regime $b\gtrsim 1/\Lambda$, via an ansatz
as follows. Defining
\begin{equation}
\widetilde{W}(b;Q,x_A,x_B) \equiv \sum_je_j^2 \widetilde{W}_j(b;Q,x_A,x_B),
\end{equation}
which we see from (\ref{css2}) and (\ref{css3}) is well-defined for both regimes of interest in $b$, the authors in Ref.~\cite{colsopster} use
\begin{equation}
b^* = b/\sqrt{1+b^2/b_{\text{max}}^2}
\end{equation}
where $b_{\text{max}}$ is a parameter still in the perturbative regime
where (\ref{css2}) holds, to write the joining ansatz
\begin{equation}
\widetilde{W}_j(b;Q,x_A,x_B) = \widetilde{W}_j(b^*;Q,x_A,x_B) e^{\{-\ln(Q^2/Q_0^2)g_1(b)-g_{j/A}(x_A,b)-g_{j/B}(x_B,b)\}},
\end{equation}
where the functions $g_1,\; g_{j/A},\;g_{j/B}$ have non-perturbative content and must be determined from data. With this ansatz, the authors in Ref.~\cite{colsopster} arrive at the representation{\small
\begin{align}
\frac{d\sigma}{dQ^2dydQ_T^2}&\sim \frac{4\pi^2\alpha^2}{9Q^2s}\Bigg\{\int \frac{d^2b}{(2\pi)^2} e^{i\vec{Q}_T\cdot\vec{b}}\sum_je_j^2\widetilde{W}_j(b^*;Q,x_A,x_B)e^{\tiny\{-\ln(Q^2/Q_0^2)g_1(b)-g_{j/A}(x_A,b)-g_{j/B}(x_B,b)\}} \nonumber\\
&\qquad\qquad\qquad\qquad\qquad +\; Y(Q_T;Q,x_A,x_B)\Bigg\}.
\label{css4} 
\end{align}}
We point-out for completeness and illustration that in R{\scriptsize ES}B{\scriptsize OS} in Refs.~\cite{resbos2,resbos3} three realizations of the 
non-perturbative functions in (\ref{css4}) were considered as follows:
\begin{align}
& \ln(Q^2/Q_0^2)g_1(b)+g_{j/A}(x_A,b)+g_{j/B}(x_B,b)\nonumber\\
&\qquad\qquad =\begin{cases}[g_1+g_2\ln(Q/(2Q_0))]b^2,\; \text{DWS~\cite{dws}}\\
[g_1+g_2\ln(Q/(2Q_0))]b^2+g_1g_3\ln(100x_Ax_B)b,\; \text{LY~\cite{resbos2}}\\
[g_1+g_2\ln(Q/(2Q_0))+g_1g_3\ln(100x_Ax_B)]b^2,\; \text{BLNY~\cite{resbos3}}.
                                                  \end{cases}
\label{css5}
\end{align}
We observe that the $g_i$ are parameters on the RHS of (\ref{css5}) and that the best fit to the data considered in Ref.~\cite{resbos3}
was obtained from the BLNY parametrization with the values 
\begin{equation}
g_1=0.21^{+0.01}_{-0.01} \text{GeV}^2,\; g_2=0.68^{+0.01}_{-0.02}\text{GeV}^2,\; g_3=-0.6^{+0.05}_{-0.04}
\label{css6}
\end{equation} 
when $b_{\text{max}}=0.5 \text{GeV}^{-1}$ and $Q_0= 1.6 \text{GeV}$.
\par
On the question of the physical precision of (\ref{css4}), we first observe that, with the errors shown on the constants $g_i$
for the best BLNY parmetrization, the $-0.02$ error on $g_2$ represents already a 1.5\% uncertainty at the saddle point position for the integration over $b$ in the respective $Z$ production analogue of
(\ref{css1}) found in Ref.~\cite{colsopster} and recapitulated below in
(\ref{css7}) -- this alone exceeds the theory error budget in precision QCD theory that we advocate here so that one would need these errors reduced considerably for them to be useful for our purposes. But, what is more important is that, on the LHS of (\ref{css5}) the $g_\ell$ which 
do not (do) multiply $\ln(Q/Q_0)$ are unspecified {\em functions} of $x_j,b\;(b)$ that are required to vanish at $b=0$ while on the RHS these dependences are simplified to second order polynomials in $b$ and either no or linear $\ln(x_Ax_B)$ dependence on $x_j$: these simplifications are generically ad hoc
and can not be considered as a rigorous platform for testing the fundamental QCD theory. They(the simplifications) do not follow from the formalism of Ref.~\cite{colsop,colsopster}. As the authors in Ref.~\cite{dws} have emphasized, the latter formalism could only be used to give a prediction for perturbative QCD in the regime where the 
details of the non-perturbative parametrization are not important. We can not emphasize this point too much.\par
On the question of the physical precision of (\ref{css4}) we additionally note that in the regime $0\le Q_T\ll Q$ it has an error of ${\cal O}(Q_T/Q)$ according to 
the authors in Ref.~\cite{colsop,colsopster}, as such terms have been dropped, order by order in perturbation theory in the construction of $\widetilde{W}_j$ -- 
we make reference again specifically to Eq.(9.1) and the equation immediately preceding it in Ref.~\cite{colsop}. For $Q=M_Z$ and $Q_T= 5$GeV, this is an error of $\cong 5.5\%$ and its twice this size at 
$Q_T= 10$GeV. This shows that the formalism in Ref.~\cite{colsop,colsopster} can not be
used for (sub-)1\% theory predictions for the heavy gauge boson $p_T$ spectrum
at the LHC. One might think that as $Q_T \rightarrow 0$, where these dropped terms
should vanish, the theory in Refs.~\cite{colsop,colsopster} would be accurate enough
for (sub-)1\% precision QCD predictions. This is not true as well because 
at $Q_T=0$, following Parisi and Petronzio in Refs.~\cite{ddt-res-hvy-bsn}, the authors in Ref.~\cite{colsopster} show that the integral over $\vec{b}$ in (\ref{css1}) is dominated by a saddle point at
\begin{equation}
b_{\text{SP}}=\frac{1}{\Lambda}\left(\frac{Q}{\Lambda}\right)^{-A^{(1)}/(A^{(1)}+\beta_1)}
\label{css7}
\end{equation}  
where we have $A=\sum_NA^{(N)}(\alpha_s/\pi)^N$ and $\beta_1=\frac{1}{12}(33-2N_F)$ where $N_F$ is the number of effective quark flavors, 
which we take as $5$ here. Thus, we have~\cite{colsop,colsopster} $ A^{(1)}=4/3$ so that
$$b_{\text{SP}}=\frac{1}{\Lambda}\left(\frac{Q}{\Lambda}\right)^{-0.41}.$$
For $Q=M_Z$, we get $b_{\text{SP}}\cong .48\text{GeV}^{-1}$ when we follow Ref.~\cite{colsopster}
and take $\Lambda=0.15\text{GeV}$. This result
for $b_{\text{SP}}$ is in the perturbative regime. Moreover, we know from Refs.~\cite{colsop,colsopster} that the error on the $Z$ production analogue
of $\widetilde{W}_j$
at this point is ${\cal O}(1/(b_{\text{SP}}M_Z))\cong 2.3\%$ -- see again Eq.(9.1) in Ref.~\cite{colsop}.
Thus, even when the perturbative regime obtains so that the result (\ref{css2})
should be reliable, the error in it is too large for use as anything but a 
guide in constructing our precision $QCD\otimes EW$ theory residuals $\hat{\tilde{\bar\beta}}_{n,m}$ 
in our master formula (\ref{subp15b}) for $\lesssim 1\%$ precision LHC physics.
The results reported in Refs.~\cite{atlas-z-pt-2} on the comparison of 
R{\scriptsize ES}B{\scriptsize OS} with recent ATLAS and Tevatron data
related to the $p_T$ spectrum in single $Z/\gamma^*$ are consistent with
our estimates on the physical precision of (\ref{css4}): R{\scriptsize ES}B{\scriptsize OS} misses the data by ~2\% for the $p_T$ near 0 and it misses the data
by $\gtrsim 5\%$ for the regime of $p_T\gtrsim 10\text{GeV}$.\par 
The non-perturbative factor in (\ref{css4}) compromises the predictive
power of the formalism in Refs.~\cite{colsop,colsopster}. The authors in Ref.~\cite{dws} have argued that for W,~Z production as considered here the prediction in
(\ref{css4}) in the regime $6\text{GeV}<Q_T<16\text{GeV}$ is insensitive to the non-perturbative parametrization in (\ref{css5}). This leaves the effects we have discussed above as the main obstacles to using the theory in Refs.~\cite{colsop,colsopster} for precision QCD physics predictions at the LHC in this latter regime when that precision is at or below 1\%.
\par
We note that in Ref.~\cite{banfi}, another version
of (\ref{css1}) is presented which has the same physical precision limits
as that realized in R{\scriptsize ES}B{\scriptsize OS} with a different 
treatment of the non-perturbative regime. As the comparison with the data shown in Refs.~\cite{atlas-z-pt-2} shows
in the perturbative regime, our estimates of this physical precision
for this regime also apply to the realization of (\ref{css1}) in Ref.~\cite{banfi}: it cannot be used for 1\% precision LHC physics studies.\par
Here, let us note that the authors in Ref.~\cite{banfi} discuss the 
uncertainty in their
results as a function of the variation 
of the (perturbative scale) values associated with their 
renormalization, factorization and resummation scales. 
For example, they estimate 
that this is at the level of 10\% near the peak of the $Q_T$ spectrum in 
the single $Z/\gamma*$ production at the LHC. These uncertainties  can in 
principle be reduced by computing the perturbative terms in their results to 
higher and higher orders. They also estimate an additional PDF error at this 
region at the level of 2\%, which in principle can be reduced by improving 
the determination of the respective PDF's. We stress that the error in the 
defining result (\ref{css1}) derived in Refs.~\cite{colsop,colsopster}
on which their results are based that we discuss here is an error of 
${\cal O}(Q_T/Q)$ that applies order by order in the perturbation 
theory -- it is separate from the scale and PDF errors discussed in 
Ref.~\cite{banfi}.\par
Indeed, in Ref.~\cite{neubrt}, the SCET approach is used to recover
(\ref{css1}) and it is shown explicitly that the improvement of the perturbative scale errors indeed occurs when higher order corrections are included in the calculation of the perturbative terms in the respective SCET realization of (\ref{css1}). We stress again that the error that we discuss here of ${\cal O}(Q_T/Q)$
due to the approximations in the defining derivation of (\ref{css1}) in Refs.~\cite{colsop,colsopster} also applies to the order-by-order results in Refs.~\cite{neubrt}. As we noted above, SCET involves for single $Z$ production at the LHC
the defining error $\sqrt{\Lambda/M_Z}\sim 5.7\%$ for the typical hadronic
transverse size $\Lambda\cong 0.3\text{GeV}$ and this is consistent with
the approximations made in Refs.~\cite{colsop,colsopster}. Such approximations
cannot be used for the 1\% precision QCD theory that we have as our goal for (\ref{subp15b}) here.\par 
In summary, this last remark is beginning to be manifest as we see in Refs.~\cite{atlas-z-pt-2} in the comparison between
the recent Tevatron and LHC data on the $Q_T$ spectra in the $Z/\gamma*$ 
production and the predictions of R{\scriptsize ES}B{\scriptsize OS} and
of Ref.~\cite{banfi}. 
Indeed, even though a new $Q_T$-related variable~\cite{phietastr} is used in
some of the comparisons, $\phi_\eta^*=\tan(\frac{1}{2}(\pi-\Delta\phi))\sin\theta^* \cong \left|\sum \frac{{p_i}_T \sin\phi_i}{Q}\right| +{\cal O}(\frac{{{p_i}_T}^2}{Q^2})$, where $\Delta\phi=\phi_1-\phi_2$ is the azimuthal angle 
between the two leptons which have transverse momenta $\vec{p_i}_T,\; i=1,2,$
and $\theta^*$ is the scattering angle of the dilepton system relative to the beam direction when one boosts to the frame along the beam direction such that the leptons are back to back, one sees that these $\phi_\eta^*$-comparisons also show the
underlying physical precision error associated with the defining formula in (\ref{css1}); as expected, the comparisons are somewhat better than the $Q_T$
spectra comparisons because this  $\phi_\eta^*$ quantity
is more inclusive -- two different values of $Q_T$ with correspondingly compensating differences in the attendant $\phi_i$ can have the same value of $\phi_\eta^*$. We want to encourage the LHC experimentalists to continue to
produce the
real $Q_T$ spectra in the $Z/\gamma*$ production because we believe, even if there are some irreducible experimental systematic errors, these spectra can be very useful in determining which theoretical approach is actually correct as we argue in the next section. In particular, given the intrinsic physical precision error in (\ref{css1}), one has to account for that error when the data are normalized to any prediction which is based on the formula therein. We would like to stress this latter point.
We await the more complete data sets accordingly.\par 
From the discussions just completed, we see that, 
in order to have a strict control on the theoretical precision
in (\ref{bscfrla}), we need both the resummation of the reduced cross section
and that of the attendant evolution of the $\{F_j\}$. We turn now
to the latter. 
\par 
More specifically, we apply the QCD restriction of the formula in (\ref{subp15b}) to the
calculation of the kernels, $P_{AB}$, in the DGLAP-CS theory itself and thereby 
get an improvement
of the IR limit of these kernels, an IR-improved DGLAP-CS theory~\cite{irdglap1,irdglap2} in which large IR effects are resummed for the kernels themselves.
The attendant new resummed kernels, $P^{\exp}_{AB}$ are given in Refs.~\cite{irdglap1,irdglap2,herwiri}. We reproduce the new kernels here for completeness:
{\small
\begin{align}
P^{\exp}_{qq}(z)&= C_F \FYFS(\gamma_q)e^{\frac{1}{2}\delta_q}\left[\frac{1+z^2}{1-z}(1-z)^{\gamma_q} -f_q(\gamma_q)\delta(1-z)\right],\nonumber\\
P^{\exp}_{Gq}(z)&= C_F \FYFS(\gamma_q)e^{\frac{1}{2}\delta_q}\frac{1+(1-z)^2}{z} z^{\gamma_q},\nonumber\\
P^{\exp}_{GG}(z)&= 2C_G \FYFS(\gamma_G)e^{\frac{1}{2}\delta_G}\{ \frac{1-z}{z}z^{\gamma_G}+\frac{z}{1-z}(1-z)^{\gamma_G}\nonumber\\
&\qquad +\frac{1}{2}(z^{1+\gamma_G}(1-z)+z(1-z)^{1+\gamma_G}) - f_G(\gamma_G) \delta(1-z)\},\nonumber\\
P^{\exp}_{qG}(z)&= \FYFS(\gamma_G)e^{\frac{1}{2}\delta_G}\frac{1}{2}\{ z^2(1-z)^{\gamma_G}+(1-z)^2z^{\gamma_G}\},
\label{dglap19}
\end{align}}
where the superscript ``$\exp$'' indicates that the kernel has been resummed as
predicted by Eq.\ (\ref{subp15b}) when it is restricted to QCD alone. Here
$C_F$($C_G$) is the quadratic Casimir invariant for the quark(gluon) color representation respectively, and  
the YFS~\cite{yfs} infrared factor 
is given by $$\FYFS(a)=e^{-C_Ea}/\Gamma(1+a)$$ where $\Gamma(w)$ is Euler's gamma function and $C_E=0.57721566...$ is Euler's constant. 
The definitions of the respective resummation functions $\gamma_A,\delta_A,f_A, A=q,G$ are as follows~\cite{irdglap1,irdglap2}
\footnote{The improvement in Eq.\ (\ref{dglap19}) 
should be distinguished from the 
resummation in parton density evolution for the ``$z\rightarrow 0$'' 
Regge regime -- see for example Refs.~\cite{ermlv,guido}. This
latter improvement must also be taken into account 
for precision LHC predictions.}:
\begin{align}
\gamma_q &= C_F\frac{\alpha_s}{\pi}t=\frac{4C_F}{\beta_0}, \qquad \qquad
\delta_q =\frac{\gamma_q}{2}+\frac{\alpha_sC_F}{\pi}(\frac{\pi^2}{3}-\frac{1}{2}),\nonumber\\
f_q(\gamma_q)&=\frac{2}{\gamma_q}-\frac{2}{\gamma_q+1}+\frac{1}{\gamma_q+2},\nonumber\\
\gamma_G &= C_G\frac{\alpha_s}{\pi}t=\frac{4C_G}{\beta_0}, \qquad \qquad
\delta_G =\frac{\gamma_G}{2}+\frac{\alpha_sC_G}{\pi}(\frac{\pi^2}{3}-\frac{1}{2}),\nonumber\\
f_G(\gamma_G)&=\frac{n_f}{6C_G \FYFS(\gamma_G)}{e^{-\frac{1}{2}\delta_G}}+
\frac{2}{\gamma_G(1+\gamma_G)(2+\gamma_G)}+\frac{1}{(1+\gamma_G)(2+\gamma_G)}\nonumber\\
&\qquad +\frac{1}{2(3+\gamma_G)(4+\gamma_G)}+\frac{1}{(2+\gamma_G)(3+\gamma_G)(4+\gamma_G)},
\label{resfn1}
\end{align}
where $\Gamma(w)$ is Euler's gamma function and $C$ is Euler's constant.
We use a one-loop formula for $\alpha_s(Q)$ in (\ref{resfn1}), so that
\[\beta_0=11-\frac{2}{3}n_f\equiv 4\beta_1,\] where $n_f\equiv N_F$ is the number of
active quark flavors. 
These new kernels provide us with a new resummed scheme for the parton density functions (PDF's) and the reduced cross section with the same value of $\sigma$ in (\ref{bscfrla}): 
\begin{equation}
\begin{split}
F_j,\; \hat\sigma &\rightarrow F'_j,\; \hat\sigma'\; \text{for}\\
P_{Gq}(z)&\rightarrow P^{\exp}_{Gq}(z), \text{etc.}.
\end{split}
\label{newscheme1}
\end{equation}
This new scheme has improved MC stability
as discussed in Refs.~\cite{herwiri} -- in the attendant parton shower MC based on the new kernels there is no need 
for an IR cut-off `$k_0$'
parameter.
It is important to note that, while the degrees of freedom
below the IR cut-offs in the usual showers are dropped in those showers,
in the showers in HERWIRI1.031,
as one can see from (\ref{subp15b}), these degrees of freedom are included in the calculation and are integrated over in the process of generating the Gribov-Lipatov exponents $\gamma_A$ in (\ref{dglap19}). We note also that {\em the new kernels
agree with the usual kernels at ${\cal O}(\alpha_s)$ as the differences between them start in ${\cal O}(\alpha_s^2)$}. This means that the NLO matching formulas
in the MC@NLO and POWHEG frameworks apply directly 
to the new kernels for the realization of exact
NLO ME/shower matching. 
\par
In the interest of pedagogy, in Fig.~1 we illustrate the basic physical idea,
already discussed by Bloch and Nordsieck~\cite{bn1}, which underlies the new kernels: 
\begin{figure}[h]
\begin{center}
\epsfig{file=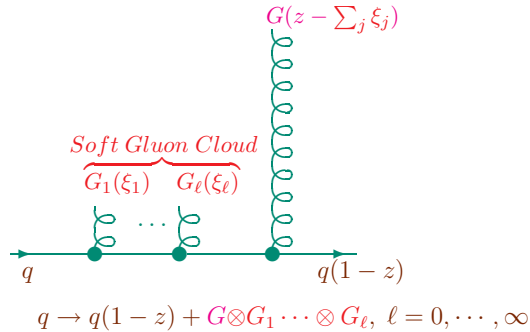,width=70mm}
\end{center}
\label{fig-bn-1}
\caption{Bloch-Nordsieck soft quanta for an accelerated charge.}
\end{figure}
an accelerated charge generates a coherent state of very soft massless quanta of the respective gauge field so that one cannot know which of the infinity of possible states
one has made in the splitting process $q(1)\rightarrow q(1-z)+G\otimes G_1\cdots\otimes G_\ell,\; \ell=0,\cdots,\infty$ illustrated in Fig.~1.
In the new kernels this effect is taken into account by resumming the 
terms ${\cal O}\left((\alpha_s \ln(\frac{q^2}{\Lambda^2})\ln(1-z))^n\right)$
when $z\rightarrow 1$ is the IR limit. From (\ref{newscheme1}) and (\ref{bscfrla}), we see that when the usual kernels are used these terms
are generated order-by-order in the solution for the cross section
$\sigma$ in (\ref{bscfrla}) so that our 
resumming them enhances the convergence of the 
representation in (\ref{bscfrla}) for a given order of exactness in the
input perturbative components therein.  In the next Section, we illustrate
this last remark in the context of the comparison of recent LHC data to
NLO parton shower/matrix element matched predictions.\par

\section{Interplay of IR-Improved DGLAP-CS Theory and NLO Shower/ME Precision: Comparison with LHC Data}
In the new MC HERWIRI1.031~\cite{herwiri} we have the first realization of the new IR-improved kernels in the HERWIG6.5~\cite{herwig} environment. Here, 
using recent LHC data as our baseline,
we compare it with HERWIG6.510, both with and without
the MC@NLO~\cite{mcatnlo} exact ${\cal O}(\alpha_s)$ correction
to illustrate the interplay between the attendant precision in NLO ME matched parton shower MC's  
and the new IR-improvement for the kernels.\par
\begin{figure}[h]
\begin{center}
\setlength{\unitlength}{0.1mm}
\begin{picture}(1600, 930)
\put( 370, 770){\makebox(0,0)[cb]{\bf (a)} }
\put(1240, 770){\makebox(0,0)[cb]{\bf (b)} }
\put(   -50, 0){\makebox(0,0)[lb]{\includegraphics[width=80mm]{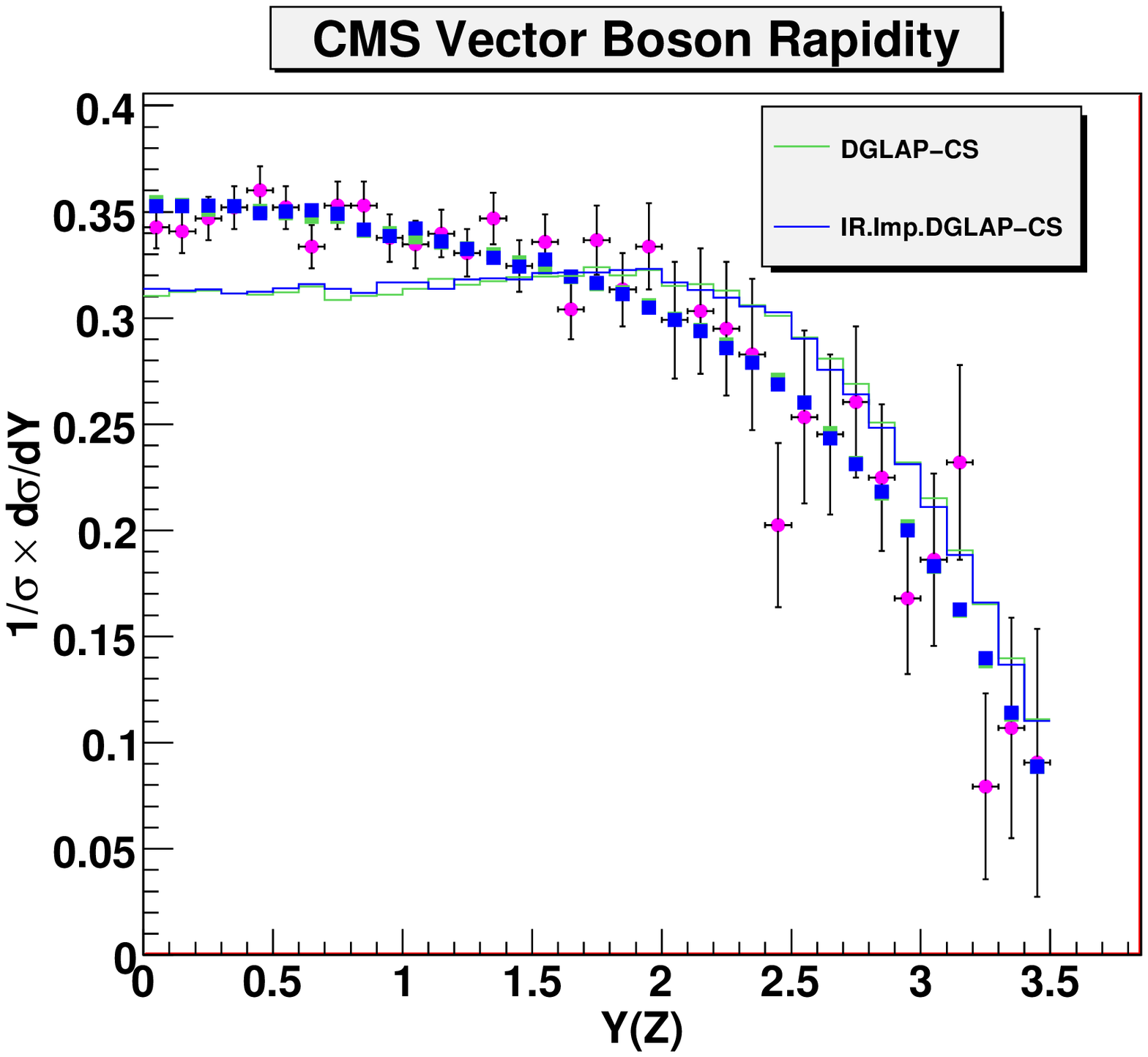}}}
\put( 830, 0){\makebox(0,0)[lb]{\includegraphics[width=80mm]{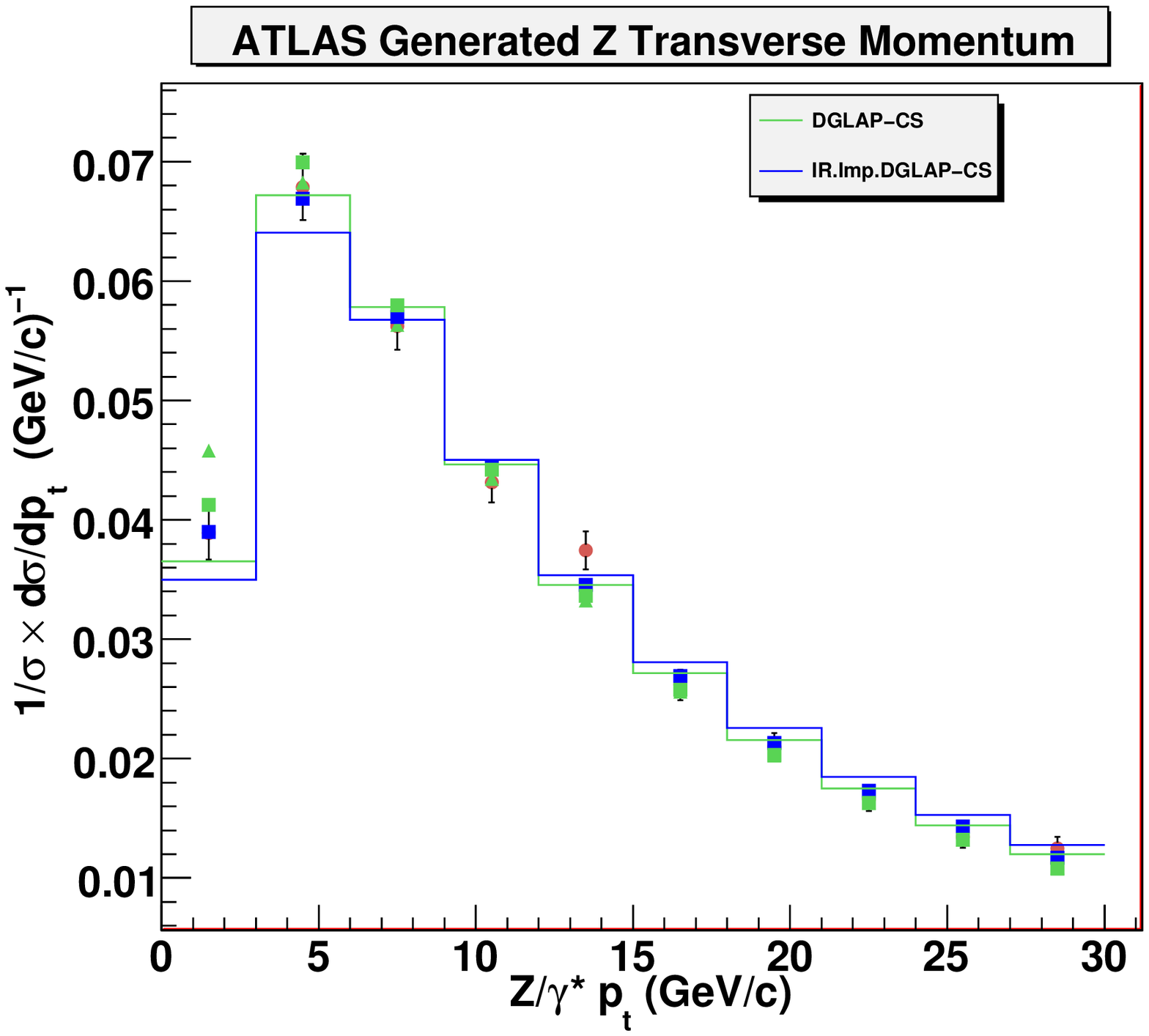}}}
\end{picture}
\end{center}
\caption{\baselineskip=8pt Comparison with LHC data: (a), CMS rapidity data on
($Z/\gamma^*$) production to $e^+e^-,\;\mu^+\mu^-$ pairs, the circular dots are the data, the green(blue) lines are HERWIG6.510(HERWIRI1.031); 
(b), ATLAS $p_T$ spectrum data on ($Z/\gamma^*$) production to (bare) $e^+e^-$ pairs,
the circular dots are the data, the blue(green) lines are HERWIRI1.031(HERWIG6.510). In both (a) and (b) the blue(green) squares are MC@NLO/HERWIRI1.031(HERWIG6.510($\rm{PTRMS}=2.2$GeV)). In (b), the green triangles are MC@NLO/HERWIG6.510($\rm{PTRMS}=$0). These are otherwise untuned theoretical results. 
}
\label{fig2-nlo-iri}
\end{figure}

More specifically, in Fig.~\ref{fig2-nlo-iri} in panel (a) we show for the single $Z/\gamma*$ production at the LHC 
the comparison between the CMS rapidity data~\cite{cmsrap} and the MC theory predictions and in panel
(b) in the same figure we show the analogous comparison with the ATLAS $P_T$ data. Here, the rapidity data  are the combined $e^+e^--\mu^-\mu^+$ results and the $p_T$ data are those for the bare $e^+e^-$ case; for, the theoretical
framework of our simulations corresponds to these data. We do not as yet have complete realization of all the corrections involved in the other ATLAS data in Ref.~\cite{atlaspt}. 
These results are better appreciated if they are considered from the perspective of our analysis in Ref.~\cite{herwiri} of the FNAL data on the single $Z/\gamma^*$ production in 
$\text{p}\bar{\text{p}}$ collisions at 1.96 TeV.\par
More precisely, we direct the reader to the results in Fig.~11 of the second paper in Ref.~\cite{herwiri}. In that figure, we showed that, when the intrinsic rms $p_T$ parameter $\rm{PTRMS}$ is set to 0 in HERWIG6.5, the MC@NLO/HERWIG6.510 simulations give a good fit to the CDF rapidity distribution data~\cite{galea} therein but they do not give a satisfactory fit to the D0 $p_T$ distribution data~\cite{d0pt} therein. In contrast, the corresponding simulations for MC@NLO/HERWIRI1.031 give good fits to both sets of data with the $\rm{PTRMS} =0$. Here $\rm{PTRMS}$ corresponds to rms value for an intrinsic Gaussian distribution in $p_T$. The authors of HERWIG~\cite{mike2} already have observed that, to get good fits to both sets of data, one may set $\rm{PTRMS}\cong 2$ GeV. Accordingly, in analyzing the new LHC data, we have set $\rm{PTRMS}=2.2$GeV in our HERWIG6.510 simulations while we continue to set PRTMS=0 in our HERWIRI simulations.
\par
We turn now with this perspective to the results in Fig.~\ref{fig2-nlo-iri}, where we see a confirmation of the finding of the HERWIG authors. One needs to set $\rm{PTRMS}\cong 2 \text{GeV}$~\cite{skands} in the MC@NLO/HERWIG6510 simulations
to get a good fit to both the CMS rapidity data and the ATLAS $p_T$ data. We again see that the MC@NLO/HERWIRI1.031 simulations with $\rm{PTRMS}=0$ at LHC give a good fit to the data for both the rapidity and the $p_T$ spectra. 
Quantitatively, we use the  $\chi^2/\text{d.o.f.}$ as a measure of the goodness
of the respective fits. From the results in Fig.~\ref{fig2-nlo-iri} we compute
that
the $\chi^2/\text{d.o.f.}$ for the rapidity data and the $\chi^2/\text{d.o.f.}$ for the
$p_T$ data are (.72,.72)((.70,1.37)) for the 
MC@NLO/HERWIRI1.031(MC@NLO/HERWIG
6510($\rm{PTRMS}$=2.2GeV)) simulations.\\ The corresponding results are (.70,2.23) for the
 MC@NLO/HERWIG6510($\rm{PTRMS}$=0) simulations.
\par 
 To reproduce the LHC data on the $p_T$ distribution of the 
$Z/\gamma^*$ in the pp collision the usual DGLAP-CS kernels require the introduction of a {\em hard}
intrinsic Gaussian distribution in $p_T$  inside the proton whereas the IR-improved kernels give in fact a better fit to the data without the introduction of such a hard intrinsic component to the motion of the proton's constituents. 
The {\em hardness}
of this intrinsic $\rm{PTRMS}$ is the issue as this quality of it is
entirely ad hoc; it is in disagreement with the results of all successful models of the proton wave-function~\cite{pwvfn},
wherein the scale of the corresponding intrinsic $\rm{PTRMS}$ is found to be 
$\lesssim 0.4$GeV. More significantly, it contradicts the well-known experimental observation of {\em precocious} Bjorken scaling~\cite{scaling,bj1}; for, the famous
SLAC-MIT experiments on the deep inelastic electron-proton scattering process show that Bjorken scaling occurs already at $Q^2=1_+$ GeV$^2$
for $Q^2=-q^2$ with q the 4-momentum transfer from the electron to the proton.
If the proton constituents really had a Gaussian intrinsic $p_T$ distribution with $\rm{PTRMS}\cong 2$GeV, these pioneering SLAC-MIT observations 
would not be possible. What we advocate now is that the ad hoc ``hardness''
of the $\rm{PTRMS}\cong 2.2$GeV value is really just a phenomenological representation of the more fundamental dynamics described by the IR-improved DGLAP-CS theory. This raises the following question: ``Is possible
to tell the difference between the two representations of the data in 
Fig.~\ref{fig2-nlo-iri}?''\par
\begin{figure}[h]
\begin{center}
\epsfig{file=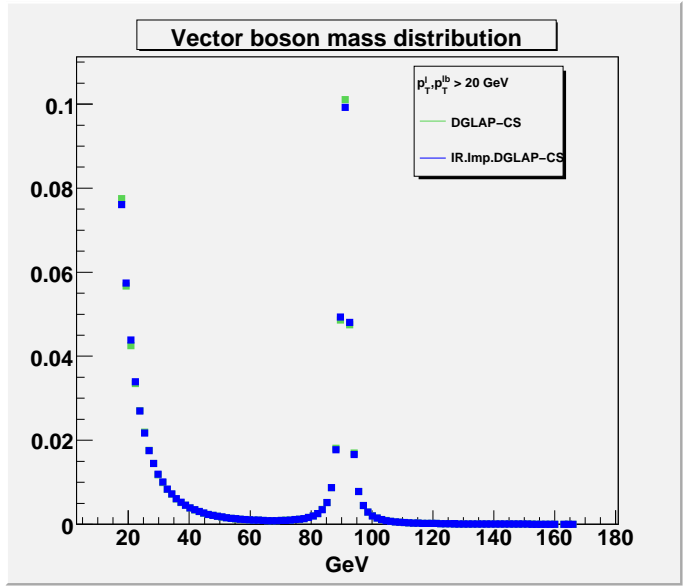,width=90mm}
\end{center}
\label{fig3-nlo-iri}
\caption{Normalized vector boson mass spectrum at the LHC for $p_T(\text{lepton}) >20$ GeV.}
\end{figure}

One expects physically that more detailed observations should be able to distinguish the two representations of the data in 
Fig.~\ref{fig2-nlo-iri}. In this connection, in Fig.~\ref{fig3-nlo-iri} we show for the $Z/\gamma^*$ mass spectrum the MC@NLO/HERWIRI1.031(blue squares) and MC@NLO/HER-\\
WIG6510($\rm{PTRMS}$=2.2GeV) (green squares) predictions when the decay lepton pairs  
satisfy the LHC type requirement that their transverse momenta $\{p^\ell_T, p^{\bar\ell}_T\}$ exceed $20$ GeV.
From the results in Fig.~\ref{fig3-nlo-iri}, wherein the peaks differ by 2.2\% for example, we see that the high precision data such as the LHC ATLAS and CMS
experiments will have
(each already has over $5\times 10^6$ lepton pairs) would allow one to distinguish between the two sets
of theoretical predictions.\par
Continuing in this direction with the discussion, we see that the main differences between the three predictions in Fig.~\ref{fig2-nlo-iri} (b) occur in the regime below $10$ GeV/c. To better probe this latter regime, we
make a more detailed snap-shot of it in which we plot in Fig.~4 the three 
respective featured theory predictions with the finer binning of $0.5$GeV/c instead of the $3.0$GeV/c binning
used in Fig.~\ref{fig2-nlo-iri} (b).  
\begin{figure}[h]
\begin{center}
\epsfig{file=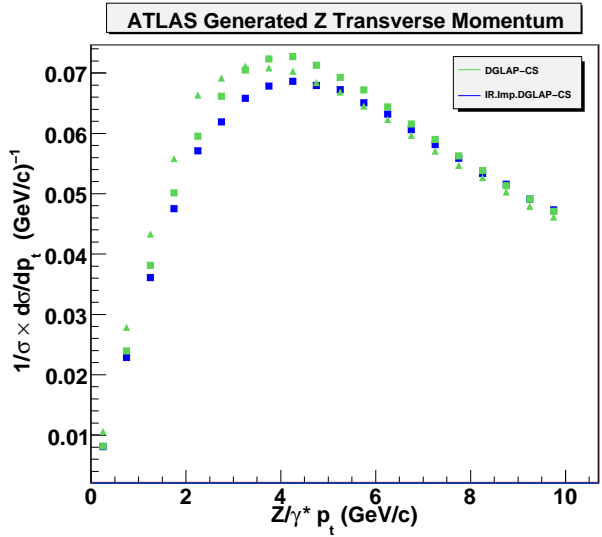,width=90mm}
\end{center}
\label{fig4-nlo-iri}
\caption{Normalized vector boson $p_T$ spectrum at the LHC 
for the ATLAS cuts as exhibited in Fig.~\protect{\ref{fig2-nlo-iri}} 
for the same conventions on the notation for the theoretical results
with the vector boson $p_T < 10$ GeV to illustrate the differences between
the three predictions.}
\end{figure}
The results in Fig.~4 show that the three theoretical predictions have significant differences in the shapes that are testable
with the precise data that will be available
to the ATLAS and CMS experiments. We would again encourage experimentalists
to pursue the measurements of both $p_T$ and $\phi_\eta^*$ spectra as these
will both be very useful in establishing the correct theoretical approach
to the respective LHC observations.
We will pursue elsewhere~\cite{elswh} other such detailed observations that
may also reveal 
the differences between the two descriptions of parton shower physics.
In this connection especially,
we continue to await the release of the entire data sets from ATLAS and CMS.\par
\section{Conclusions}
What we have shown is the following. The realization of IR-improved DGLAP-CS theory 
in HERWIRI1.031, when used in the MC@NLO/HERWIRI1.031 exact ${\cal O}(\alpha_s)$ ME matched parton shower framework,
affords one the opportunity to explain, on an event-by-event basis, both the rapidity and the $p_T$ spectra of the $Z/\gamma^*$ in pp collisions
in the recent LHC data from CMS and ATLAS, respectively, without the need of an
unexpectedly hard intrinsic Gaussian $p_T$ distribution with rms value of $\rm{PTRMS}\cong 2$ GeV in the proton's wave function. Our view is that this can be interpreted as providing a rigorous basis for the phenomenological correctness 
of such unexpectedly hard distributions insofar as describing these data using the usual unimproved DGLAP-CS showers is concerned. Accordingly, we 
have proposed 
that comparison of other distributions such as the invariant mass distribution 
with the appropriate cuts and the more detailed $Z/\gamma^*$ $p_T$ 
spectra in the regime below $10.0$GeV be used to
differentiate between these phenomenological 
representations of parton shower physics
in MC@NLO/HERWIG6510 and the fundamental description of the parton shower physics in MC@NLO/HERWIRI1.031. We have further emphasized that the precociousness of Bjorken scaling argues against the fundamental correctness 
of the {\em hard} scale intrinsic $p_T$ ansatz with the unexpectedly hard value of $\rm{PTRMS}\cong 2$ GeV, as do the successful models~\cite{pwvfn} of the proton's wave function,
which would predict this value to be $\lesssim 0.4$GeV. As an added bonus,
we have pointed-out that the fundamental description in MC@NLO/HERWIRI1.031 can be systematically improved to the NNLO parton shower/ME matched level -- a level which we anticipate is a key ingredient in achieving the (sub-)1\% precision tag for such processes as single heavy gauge boson production at the LHC.\par 
Evidently, relative to what one could achieve
from the fundamental representation of the corresponding physics via IR-improved DGLAP-CS theory as it is realized in HERWIRI1.031 when employed in
MC@NLO/HERWIRI1.031 simulations, the use of ad hoc hard scales in models would compromise
any discussion of the attendant theoretical precision. We are pursuing additional cross checks
of the MC@NLO/HERWIRI1.031 simulations against the LHC data. \par
In our discussion, we have also spent some amount of time discussing 
alternative approaches to the type of resummation
embodied in the IR-improved DGLAP-CS theory; for, some of these approaches
are in wide use.
What we conclude is that the physical precisions of these approaches(see Sect. 2), which are based on Refs.~\cite{stercattrent1,scet1,colsop,colsopster}, are above the 1\% precision tag that we aspire, even though there is no contradiction between our exact approach and these more approximate methods.\par
\par
In closing, two of us (A.M. and B.F.L.W.)
thank Prof. Ignatios Antoniadis for the support and kind 
hospitality of the CERN TH Unit while part of this work was completed.\par

\end{document}